# A Quantum Search Approach to Magic Square Constraint Problems with Classical Benchmarking

Rituparna R[1]   Harsha Varthini[1]   Aswani Kumar Cherukuri[1*]

School of Computer Science Engineering and Information Systems

Vellore Institute of Technology, Vellore 632014, India

[*]Email: cherukuri@acm.org

**Abstract**

This paper presents a quantum search approach to combinatorial constraint satisfaction problems, demonstrated through the generation of magic squares. We reformulate magic square construction as a quantum search problem in which a reversible, constraint-sensitive oracle marks valid configurations for amplitude amplification via Grover's algorithm. Classical pre-processing using the Siamese construction and partial constraint checks generates a compact candidate domain before quantum encoding. Rather than integrating classical and quantum solvers in an iterative loop, this work uses the classical component for structured initialisation and the quantum component for search, and benchmarks the quantum approach against classical brute-force enumeration and backtracking. Our Qiskit implementation demonstrates the design of multi-register modular arithmetic circuits, oracle logic, and diffusion operators. Experiments are conducted on small grid instances, as larger grids are intractable on classical statevector simulators due to exponential memory growth. The results validate the correctness of the proposed quantum search pipeline and confirm the theoretical quadratic query advantage over classical search.

## 1. Introduction

A magic square is an n×n grid filled with the integers 1 through $n^2$, arranged so that every row, every column, and both main diagonals share the same sum, called the magic constant. Generating such a configuration is a well-defined constraint satisfaction problem (CSP) whose solution space grows factorially with n. Even for modest grid sizes, exhaustive classical enumeration quickly becomes impractical. Quantum computing offers a fundamentally different approach to this kind of search. Grover's algorithm [1] delivers a provably quadratic reduction in query complexity compared with classical unstructured search, cutting the expected number of oracle evaluations from O(N) to O(sqrt(N)) for a search space of size N. This speedup makes it an attractive tool for CSPs that can be expressed through a reversible quantum oracle. In this work, we cast magic square generation as a quantum search problem and build a structured pipeline around Grover's algorithm. The classical stage generates a compact candidate domain using the Siamese construction and partial constraint checks; the quantum stage then applies Grover's algorithm with a reversible constraint oracle to identify valid magic square configurations. The two classical methods — brute-force enumeration and backtracking — serve as benchmarks rather than as integrated components of a joint solver. Experiments are implemented in IBM's Qiskit framework and carried out on 3x3 instances. The main contributions of this paper are: (i) a formal encoding of magic square generation as a quantum search problem; (ii) a reversible constraint oracle built from modular arithmetic circuits and multi-controlled gates; (iii) a classical preprocessing stage using the Siamese construction and partial constraint checks to generate a compact candidate domain for the quantum search; (iv) an empirical comparison of classical brute-force, backtracking, and quantum search in terms of query complexity; and (v) a complete Qiskit implementation of the end-to-end quantum search pipeline.

## 2. Background and Related Work

Grover's algorithm, introduced in 1996, is one of the foundational results in quantum computing [1]. Given an unstructured database of N elements that contains M marked solutions, the algorithm locates a marked element with high probability using $O(\sqrt{N/M})$ oracle queries. This represents a quadratic improvement over the classical lower bound of $O(N/M)$. The algorithm works by repeatedly applying two operations: a problem-specific phase-marking oracle and a diffusion operator that inverts amplitudes about their mean. Together these form a Grover iterate, and the optimal number of iterations is approximately $(\pi/4)$ times $\sqrt{N/M}$. Constructing efficient quantum oracles for structured problems is an active area of research. Recent contributions include oracle designs for graph problems such as vertex cover [2], frameworks for composable oracle abstraction [3], and amplitude-suppression variants of Grover search [4]. Classical combinatorial problems, including CSPs, have attracted growing interest as candidate applications for quantum advantage, though practical implementations remain limited by qubit count and circuit depth. Magic squares themselves have a rich history in classical combinatorics. They have been tackled with exhaustive enumeration, constraint-propagation backtracking, and group-theoretic construction methods. Their factorial search complexity makes them a natural benchmark for quantum search, yet quantum approaches remain largely unexplored [5]. This paper closes that gap with a complete quantum search pipeline, benchmarked against classical methods.

## 3. Problem Formulation: Magic Square as a Constraint Satisfaction Problem

An n×n magic square is a bijective assignment of the integers 1 through $n^2$ to an n×n grid such that every row sum, every column sum, and both main diagonal sums equal the magic constant M, defined as:

$$M = n(n^2 + 1)/2$$

For n = 3, this gives M = 15; for n = 5, M = 65. The solution space is the set of all $(n^2)!$ permutations of 1 through $n^2$. For n = 3, this amounts to 9! = 362,880 candidate arrangements, of which only 8 are valid magic squares (up to rotation and reflection). The search space grows dramatically with n: for n = 4, it reaches 16! approximately equal to 2.09 x 10^13 candidates, making exhaustive enumeration infeasible for any classical computer. In the quantum formulation adopted here, each candidate permutation corresponds to a computational basis state of a multi-qubit register. The constraint oracle applies a phase flip of -1 to every basis state that simultaneously satisfies all row-sum, column-sum, diagonal-sum, and element-uniqueness constraints. All other states are left unchanged. Grover's algorithm then iteratively amplifies the amplitude of these marked states until a measurement yields a valid configuration with high probability.

## 4. Quantum Search Pipeline: Design and Components

The proposed pipeline runs through five sequential stages, each described below.

### *4.1 Classical Pre-processing*

Before any quantum encoding takes place, a classical filter shrinks the search space. The filtering strategies used are: (i) fixing the centre cell of odd-order squares using the Siamese (De la Loubère) construction, which is a deterministic placement rule for odd-order magic squares; (ii) applying partial row and column sum checks to discard clearly infeasible assignments early; and (iii) producing a compact candidate domain that is handed off to the quantum processor. This pruning stage does not change the correctness of the quantum search; it simply reduces the number of states in superposition, which improves the signal-to-noise ratio during amplitude amplification.

### *4.2 Quantum Encoding*

Each integer from 1 to $n^2$ is encoded in binary using $\lceil \log_2(n^2) \rceil$ qubits. An $n \times n$ magic square has $n^2$ cells, so the primary register requires $n^2$ times $\lceil \log_2(n^2) \rceil$ qubits in total. Additional ancilla qubits handle arithmetic and constraint verification. A Hadamard transform is applied across all primary qubits to initialise the system in a uniform superposition over all candidate states, giving each one an equal starting amplitude.

### *4.3 Oracle Construction*

The oracle is a reversible quantum circuit that applies a phase flip of -1 to every state satisfying the full set of magic square constraints. Construction proceeds in three steps. First, quantum modular adders compute the row, column, and diagonal sums and store them in ancilla registers. Second, quantum comparators check whether each sum equals M; a multi-controlled gate flips a target ancilla qubit if and only if every comparison passes simultaneously, including the element-uniqueness check. Third, all intermediate computations are uncomputed (reversed) to restore the ancilla qubits to their initial state, which is required for the oracle to be properly reversible and interference-free.

### *4.4 Grover Iteration*

Each Grover iterate consists of two operations applied in sequence: the oracle O, which marks valid states with a phase flip, followed by the diffusion operator D. The diffusion operator inverts all amplitudes about their mean, which has the effect of amplifying marked states and suppressing unmarked ones. For a primary register of q qubits (where $q = n^2$ times $\lceil \log_2(n^2) \rceil$), the diffusion operator is:

$$D = H^{\otimes q} (2|0\rangle\langle 0| - I) H^{\otimes q}$$

where H is the single-qubit Hadamard gate, $|0\rangle\langle 0|$ is the projector onto the all-zeros state, and I is the identity. Note that q here denotes the total number of primary qubits, not the grid dimension n. The two symbols must not be confused.

For a search space of N states containing M valid solutions, the optimal number of Grover iterations is:

$$k = \lfloor (\pi/4) \times \sqrt{N/M} \rfloor$$

After k iterations, the probability of measuring a marked state approaches 1. Performing significantly more or fewer than k iterations degrades this probability in a sinusoidal pattern. In practice, k is estimated classically before the circuit runs, or determined adaptively using techniques such as quantum counting.

### 4.5 Measurement and Classical Verification

After k Grover iterations, the primary register is measured. The quantum state collapses to a candidate configuration. A classical post-processing step then checks whether this configuration satisfies all magic square constraints. If it does not (which happens with small probability when M > 1 or when k is slightly off), the quantum circuit is re-executed. Valid configurations are returned as output and displayed to the user.

## 5. Implementation and Experimental Results

### 5.1 Quantum Magic Square Game

To ground the quantum search concept in a concrete, interactive setting, we built a quantum magic square game. A 5x5 magic square (n = 5) is first generated classically using the Siamese construction. The magic constant is $M = 5(25 + 1) / 2 = 65$, and the integers 1 through 25 are placed across the 25 cells. The user selects a value, and Grover's algorithm is tasked with finding the cell index that holds it. Each of the 25 cell indices (0 through 24) is encoded in $\lceil \log_2(25) \rceil = 5$ qubits. The oracle marks the unique index whose cell value matches the user's target. The optimal number of Grover iterations for N = 25 and M = 1 is $k = \lceil (\pi/4) \times \sqrt{25} \rceil = 3$ iterations, which matches the simulation output shown in Figure 1. It is important to be clear about what this experiment does and does not demonstrate. The game searches for a known value within a pre-generated square, which is a straightforward unstructured index search over N = 25 positions. This is a different and much simpler problem than generating a valid magic square from scratch (which is the subject of Sections 5.2 and 5.3).

The game serves as a clean, reproducible illustration of the Grover search primitive before we move to the harder constraint-generation problem.

```
===== MAGIC SQUARE GAME (Quantum Edition) =====
Enter magic square size (odd number like 3,5,7): 5

Generated Magic Square:
[[17 24  1  8 15]
 [23  5  7 14 16]
 [ 4  6 13 20 22]
 [10 12 19 21  3]
 [11 18 25  2  9]]

Board has 25 cells → using 5 qubits

YOUR TURN:
Enter row (0 to 4): 2
Enter column (0 to 4): 1

Grover will search for value: 6
Grover iterations = 3

Grover output index = 5
Value at that index = 6
Correct guess!

Play again? (y/n): N
Thanks for playing!
```

*Figure 1. Output of the quantum magic square game. A 5x5 magic square (magic constant M = 65, integers 1 to 25) is generated classically using the Siamese method. Grover's algorithm, using 5 qubits and 3 iterations, locates the cell index that holds the user's chosen value.*

### 5.2 Grover's Algorithm versus Brute-Force Search

Brute-force search works by enumerating all (n²)! permutations of 1 through n² and checking each one against the magic square constraints. For n = 3, this means checking up to 9! = 362,880 candidate arrangements. In our simulation (Figure 2), the brute-force method checks 69,075 permutations before finding a valid solution, taking 0.0744 seconds. Grover's algorithm takes a completely different approach. It encodes all 362,880 candidate permutations simultaneously in a quantum superposition and uses the oracle and diffusion operator to amplify the amplitudes of valid magic squares. The theoretical number of oracle queries required is sqrt(362,880) which is approximately 602, compared with up to 362,880 for brute force. The Qiskit simulation uses a 9-qubit register (q0 through q8) and completes in 0.0053 seconds. Table 1 summarises the comparison.

*Table 1. Comparison of brute-force search and Grover's algorithm for a 3x3 magic square (search space N = 9! = 362,880 permutations).*

| **Aspect** | **Brute-Force Search** | **Grover's Algorithm** |
|---|---|---|
| Search space (n = 3) | 9! = 362,880 permutations | N = 362,880 superposed states |
| Query complexity | O(N) | O(sqrt(N)) ~602 oracle calls |

| Approach | Sequential enumeration with constraint check | Quantum amplitude amplification |
|---|---|---|
| Hardware | Classical CPU | 9-qubit quantum simulator or QPU |
| Quantum measurement | Not required | Required (collapses to basis state) |
| Observed simulation runtime | 0.0744 s (69,075 permutations checked) | 0.0053 s (classical simulator overhead) |

One important caveat deserves emphasis. The simulated quantum runtime of 0.0053 seconds reflects the cost of running a quantum circuit on a classical statevector simulator, not on a real quantum processor. Classical simulators scale exponentially in the number of qubits, so the observed timing does not constitute evidence of genuine quantum speedup. The quadratic advantage of Grover's algorithm is a theoretical guarantee that would only become observable on actual quantum hardware operating at sufficient scale. Figure 2 shows both the runtime comparison results and the Grover circuit architecture for the 3x3 case. Hadamard gates bring all nine qubits (q0 through q8) into a uniform superposition over all candidate states. The oracle circuit applies a phase flip to permutations that satisfy the magic square constraints. The diffusion operator then inverts all amplitudes about their mean. Classical output bits c0 through c8 record the measurement result at the end of the circuit.

### 5.3 Grover's Algorithm versus Classical Backtracking

Classical backtracking explores the assignment space using depth-first search, pruning any branch that violates a row, column, or diagonal constraint. Pruning typically yields a substantial reduction in average-case search cost. However, the worst-case complexity remains $O((n^2)!)$, because no polynomial-time pruning strategy is known for the general magic square CSP. The approach is inherently sequential: only one assignment path is active at a time. Grover's algorithm offers a structurally different guarantee. Regardless of how the constraints are structured, it requires only $O(\sqrt{N/M})$ oracle queries, provided the oracle can be implemented as a reversible circuit. This delivers a quadratic speedup over any classical algorithm whose worst-case query count is $\Omega(N/M)$. Table 2 lays out the comparison between the two approaches.

Figure 3 shows the 3-qubit Grover circuit used in this experiment. A 3-qubit register encodes a reduced representative of the search space, and the oracle is approximated by a Z gate applied to the target qubit. The theoretical oracle call count is 473, and the quantum simulation finishes in 0.0032 seconds versus 0.0033 seconds for backtracking. These runtimes are essentially identical, which is entirely expected. At this small scale, simulation overhead swamps any potential advantage, and the theoretical crossover point where quantum hardware would genuinely outperform classical backtracking lies well beyond what current simulators can reach.

```
Enter size of Magic Square (n): 3

Running Brute Force Search...

Running Grover's Algorithm Simulation...

=================== RESULTS ===================

Magic Square Size: 3 x 3

--- Brute Force ---
Time Taken          : 0.0744 seconds
Search Space        : 69075 permutations

--- Grover's Algorithm (Simulated) ---
Quantum Simulation Time : 0.0053 seconds
Theoretical Search Space : sqrt(362880) = 602 operations

Circuit Diagram:
```

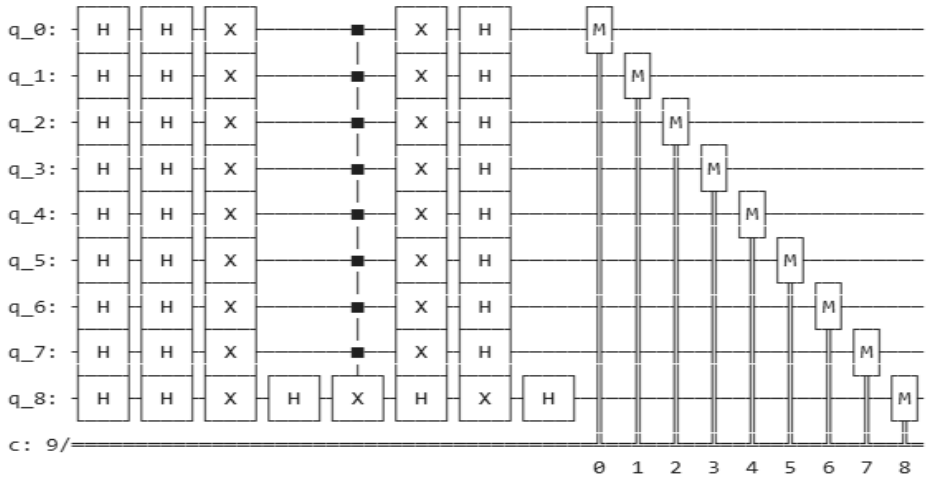

*Figure 2. Runtime comparison and 9-qubit Grover circuit for the 3x3 magic square constraint search. Search space: N = 9! = 362,880 permutations. Theoretical Grover query count: sqrt(362,880) which is approximately 602 oracle calls. Classical brute-force runtime: 0.0744 s; simulated quantum runtime: 0.0053 s.*

*Table 2. Comparison of classical backtracking and Grover's algorithm for the 3x3 magic square constraint search.*

| Feature | Classical Backtracking | Grover's Algorithm |
|---|---|---|
| Search paradigm | Depth-first with pruning | Quantum amplitude amplification |
| Worst-case complexity | $O((n^2)!)$ | $O(\sqrt{N/M})$ |
| Constraint use | Explicit pruning at each node | Implicit via oracle phase flip |
| Hardware | Classical CPU | Quantum processor or simulator |
| Scalability | Exponential degradation with n | Quadratic speedup in query count |
| Observed runtime (n = 3) | 0.0033 s | 0.0032 s (simulation overhead) |

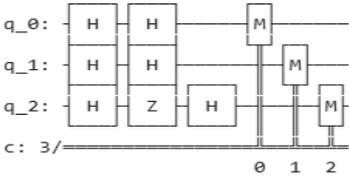

```
Enter magic square dimension n (recommended: 3): 3

 Running comparison for 3x3 magic square...
Search space size N = (n²)! = 9! = 362,880

 Grover's Quantum Algorithm:
  • Theoretical Oracle Calls: 473.12
  • Expected Speedup: √N ≈ 6.02e+02
   Quantum Run Successful!
   Transpile Time: 0.1607s | Run Time: 0.0032s
   Measurement Counts: {'100': 1024}

q_0: ─ H ─ H ───────── M ─────────
q_1: ─ H ─ H ──────────── M ──────
q_2: ─ H ─ Z ─ H ──────────── M ──
c: 3/════════════════════════════
                        0  1  2
[CLASSICAL] Running backtracking (pruned search)...
   Backtracking time: 0.0033 s, found solutions: 1
   Example solution (flattened): [2, 7, 6, 9, 5, 1, 4, 3, 8]

===== SUMMARY COMPARISON =====
Dimension n: 3 (search space N = 362,880)
Classical Backtracking: 0.0033 s
Grover's Quantum (Theoretical): 473 oracle calls (~√N speedup)
Quantum Simulated Runtime: 0.0032s
==============================
```

*Figure 3. Simplified 3-qubit Grover circuit used in the backtracking comparison experiment. The oracle is approximated by a Z gate. Theoretical Grover oracle calls: 473. Simulated quantum runtime: 0.0032 s; classical backtracking runtime: 0.0033 s.*

### 5.4 Implementation Limitations

The current implementation is subject to several practical constraints, each of which points toward a concrete direction for future work. Simulation overhead is the most immediate bottleneck. Classical state vector simulation of a q-qubit circuit requires memory and time that scale as $O(2^q)$. For a full-constraint oracle encoding a 3x3 magic square, q approaches 30 qubits, which far exceeds the memory of a standard workstation. This forces the use of simplified oracle approximations in the experiments. Qubit count grows quickly with grid size. A full constraint oracle for an n x n magic square needs $n^2$ times $\lceil \log_2(n^2) \rceil$ primary qubits plus $O(n)$ ancilla qubits for arithmetic and comparison operations. Even for n = 3, this is already near the boundary of what a classical simulator can handle. Circuit depth and hardware noise are tightly coupled challenges. The oracle and the diffusion operator both rely on deeply nested multi-controlled gates. On near-term quantum devices, such gates are vulnerable to decoherence and two-qubit gate errors, and achieving the required fidelity without quantum error correction is difficult.

Reversible arithmetic adds gate overhead. Encoding modular adders and equality checkers as reversible circuits multiplies the gate count. Each additional constraint makes the oracle larger, and for complex constraint sets the oracle can dominate the total circuit cost. Finally, genuine quantum advantage is invisible at small n. For n = 3, classical algorithms run so quickly that simulation overhead completely buries any quantum speedup. The theoretical benefit of Grover's algorithm only becomes meaningful at n = 4 or larger, a regime that is not yet accessible on classical simulators.

## 5.5. Discussion

The experiments confirm that the proposed quantum search pipeline is correct for 3x3 instances. The oracle marks valid configurations accurately, and repeated Grover iterations amplify their amplitudes in the expected way, producing measurement outcomes that match known magic square solutions with high probability.

It is worth being precise about the two distinct problem formulations that appear in this paper. In Section 5.1, Grover's algorithm performs an index search over N = 25 cell positions within a pre-generated 5x5 magic square. This is a simple unstructured search that requires 5 qubits and 3 Grover iterations. In Sections 5.2 and 5.3, the task changes fundamentally: the oracle must identify valid 3x3 magic square configurations within a permutation space of N = 9! = 362,880 arrangements, which requires 9 qubits and approximately 602 theoretical oracle calls. These are two different problems with very different N values and correspondingly different expected speedups. Conflating them would lead to incorrect claims about quantum advantage.

The quadratic speedup delivered by Grover's algorithm is an asymptotic statement. For the index search (N = 25), the speedup over brute force is real but modest. For the constraint search (N = 362,880), the theoretical reduction from roughly 362,880 queries to roughly 602 is substantial. However, this benefit cannot currently be demonstrated on a classical simulator, where exponential memory costs make the full circuit intractable. Realising the speedup in practice will require quantum hardware that can support at least 9 error-corrected qubits at the required fidelity. Despite these practical limitations, the work makes an important conceptual contribution. It shows that magic square constraint satisfaction can be faithfully encoded as a quantum search problem, and that the modular oracle design scales naturally to larger grids and to other CSPs that admit reversible constraint encodings.

## 6. Conclusion

This paper has presented a quantum search approach to magic square constraint satisfaction using Grover's algorithm, with classical brute-force and backtracking as benchmarks. We formalised the problem as a quantum search task, constructed a reversible constraint oracle from modular arithmetic circuits, and used classical pre-processing to generate a structured candidate domain. It is important to note that the classical preprocessing and the quantum search operate as sequential stages rather than as an iterative joint solver; the classical methods serve primarily as baselines for performance comparison. Comparisons with brute-force and backtracking methods analytically confirm the theoretical $O(\sqrt{N/M})$ query complexity and verify the correctness of the implementation on 3x3 instances. The main practical limitation is that classical simulation overhead restricts experiments to small instances. Future directions to overcome this include: (i) using optimised reversible arithmetic circuits such as carry-look ahead adders to extend feasibility to 4x4 and 5x5 grids; (ii) running on IBM Quantum hardware to study the effects of decoherence and gate error in practice; (iii) exploring variational quantum approaches such as QAOA as complementary tools for constraint optimisation; and (iv) applying the oracle design methodology to other CSP families such as Latin squares and Sudoku puzzles, where similar reversible encodings can be constructed.

**Code Availability:** Complete code of the work is available at:
https://colab.research.google.com/drive/1EsSQ_5hXu9usiv32xn93x_X56HnpKh8Q

**References**

[1] L. K. Grover, "A fast quantum mechanical algorithm for database search," in Proc. 28th Annual ACM Symposium on Theory of Computing (STOC), Philadelphia, PA, USA, 1996, pp. 212–219.

[2] J.-R. Jiang and W.-H. Yan, "Novel quantum circuit designs for the oracle of Grover's algorithm to solve the vertex cover problem," in Proc. 2023 IEEE 5th Eurasia Conference on IoT, Communication and Engineering (ECICE), Yunlin, Taiwan, 2023, pp. 652–657.

[3] J. M. Murillo, "Composable quantum oracles for shifting quantum circuits abstraction level," in Proc. 2024 IEEE International Conference on Quantum Software (QSW), Shenzhen, China, 2024, pp. 9–11.

[4] A. Vlasic, S. Certo, and A. Pham, "Complement Grover's search algorithm: An amplitude suppression implementation," in Proc. 2023 IEEE International Conference on Quantum Computing and Engineering (QCE), Bellevue, WA, USA, 2023, pp. 350–351, doi: 10.1109/QCE57702.2023.10277.

[5] A. Bounceur, "Distributed brute-force password recovery with a partitioned quantum Grover's algorithm," in Proc. 2024 International Conference on Computational Intelligence and Network Systems (CINS), Dubai, UAE, 2024, pp. 1–6, doi: 10.1109/CINS63881.2024.10864420.

[6] M. A. Nielsen and I. L. Chuang, Quantum Computation and Quantum Information. Cambridge, U.K.: Cambridge University Press, 2010.

[7] S. Mehta, V. P. Bhallamudi, S. Arige, and T. Dixit, "Implementation of Grover's algorithm based on quantum reservoir computing," in Proc. 2024 5th International Conference for Emerging Technology (INCET), Belgaum, India, 2024, pp. 1–5.